\documentclass{aa}
\usepackage{psfig}

\def\h2{H{\small II}}
\newcounter{qub}
\def\sbs{SBS 0335--052}
\newcommand{\hii}{H~{\sc ii}}

\newcommand{\hb}{\ifmmode {\rm H}\beta \else H$\beta$\fi}
\newcommand{\Heii}{He~{\sc ii} $\lambda$4686}
\begin{document}
%  \thesaurus{  03          % Extragalactic Astronomy
%              (11.01.1;    % Galaxies: abundances
%               11.04.2;    % Galaxies: dwarf
%               11.05.2;    % Galaxies: evolution
%               11.03.2;    % Galaxies: compact
%               11.19.3;    % Galaxies: starburst
%               11.19.5; )} % Galaxies: stellar content

\title{Deep spectroscopy of the low-metallicity blue compact 
dwarf galaxy SBS 0335--052\thanks{Data contained herein were obtained at
the W.M. Keck Observatory which is operated as a scientific partnership
among the California Institute of 
Technology, the University of California and the National Aeronautics and 
Space Administration. The Observatory was made possible by the generous 
financial support of the W.M. Keck Foundation.}}

\author{Y. I. Izotov \inst{1}
\and F. H. Chaffee \inst{2}
\and D. Schaerer \inst{3}}
\offprints{izotov@mao.kiev.ua}
\institute{      Main Astronomical Observatory,
                 Ukrainian National Academy of Sciences,
                 Golosiiv, Kyiv 03680,  Ukraine
\and
                 W. M. Keck Observatory, 65-1120 Mamalahoa Hwy., Kamuela,
                 HI 96743, USA
\and
                 Observatoire Midi-Pyr\'en\'ees, Laboratoire
d'Astrophysique, 
                 UMR 5572, 14, Av. E. Belin, F-31400 Toulouse, France}

\date{Received \hskip 2cm; Accepted}
\abstract{The results of deep long-slit spectroscopy 
of the extremely low-metallicity blue compact dwarf (BCD)
galaxy SBS 0335--052 are presented. 
Down to intensity levels of 10$^{-3 \ldots -4}$ of \hb, unprecedented 
for spectroscopy of extra-galactic giant \hii\ regions, we detect numerous 
weak permitted and forbidden nebular lines in the brightest part of the
galaxy.
With varying degrees of confidence, the detections include lines of
high-ionization ions like Fe$^{4+}$ --Fe$^{6+}$, implying very hard ionizing
radiation.
Two broad emission features, possibly from Wolf-Rayet stars, and
stellar He {\sc ii} $\lambda$4200 absorption are seen in the same region.
The large spatial extent of He {\sc ii} $\lambda$4686
emission (implying the presence of sufficient ionizing photons with energies
above 54 eV) and the spatial distribution of the electron temperature 
suggest that at least some part of the hard radiation is associated with
shocks.
Extended H$\alpha$ emission is detected over $\sim$ 6 -- 8 kpc, a much
larger 
area than in previous studies, suggesting that hot ionized gas is 
spread out far away from the central ionizing clusters. This shows that
nebular line and continuous emission can significantly modify the colours 
of these extended regions and must be taken into account in studies
of the underlying stellar population. 
}

\maketitle

\markboth {Y. I. Izotov et al.}{Deep Keck spectroscopy of the 
BCD SBS 0335--052}

\section{Introduction}

Since its discovery by Izotov et al. (1990) as a very metal-poor galaxy
the blue compact dwarf (BCD) galaxy SBS 0335--052 (SBS -- the Second
Byurakan 
Survey) has been studied extensively.
The oxygen abundance
in SBS 0335--052 is 12 + log O/H = 7.30 (Melnick, Heydari-Malayeri \& Leisy 
1992; Izotov et al. 1997, 1999) and places it after I Zw 18 as the second
most 
metal-deficient BCD. 
The properties of SBS 0335--052 as a probable young galaxy are of great 
interest for cosmology (e.g., Izotov et al. 1997).
Therefore, detailed studies of this galaxy can shed light on 
the formation and the properties of the high-redshift primeval galaxies.
Using deep spectroscopic observations we concentrate in the present paper
on two problems: (1) the origin of very hard radiation
at wavelengths shorter 228\AA. This radiation is indicated by the presence
of the strong He {\sc ii} $\lambda$4686 emission line (Izotov et al. 1997)
and implies that some other weak emission lines of high-ionization species
could be present in the spectrum of SBS 0335--052. 
(2) the analysis of the properties of 
extended emission around SBS 0335--052. This problem is of great 
importance in understanding the evolutionary status of SBS 0335--052.
In particular, we aim to understand how important gaseous emission is
in SBS 0335--052 and how far it extends from the ionizing clusters.
%Is extended emission stellar in origin, produced presumably by the evolved 
%stellar population or is gaseous in nature?

%*********************************************************
%*********************************************************
%  Fig.1 - the spectrum with line identifications
%*********************************************************
%*********************************************************

\begin{figure}
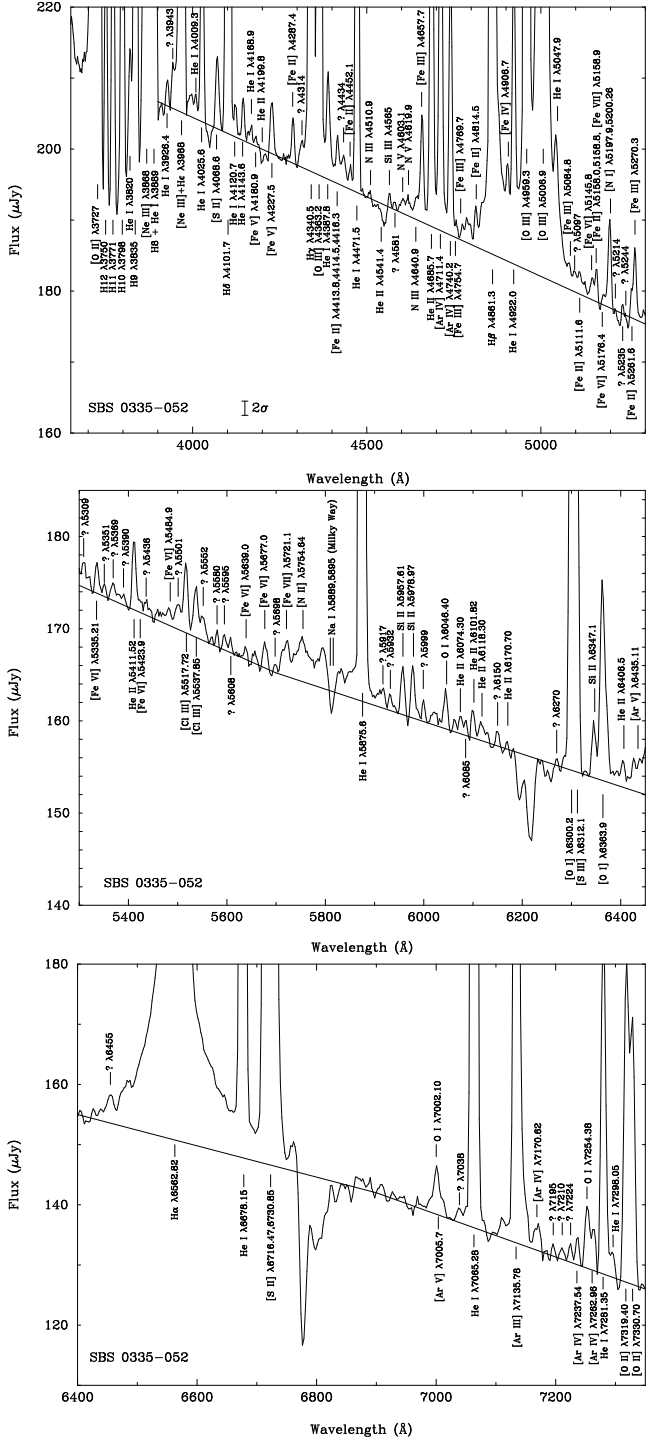
%[hbtp]
    \hspace*{0.0cm}\psfig{figure=Dh013_f1a.eps,angle=270,width=8.5cm}
    \hspace*{0.0cm}\psfig{figure=Dh013_f1b.eps,angle=270,width=8.5cm}
    \hspace*{0.0cm}\psfig{figure=Dh013_f1c.eps,angle=270,width=8.5cm}
    \caption{The spectrum of the brightest region with line
identifications. On each panel a continuum is drawn by hand to 
guide the eye. 
%The blue part of the spectrum containing only bright
%[O {\sc ii}], [Ne {\sc iii}], He {\sc i} and hydrogen emission lines
%is not shown.
2$\sigma$ error in the continuum placement is shown in the upper panel.}
    \label{fig:weaklines}
\end{figure}

%*********************************************************
%*********************************************************

%*********************************************************
%*********************************************************
%  Fig.2 - rel.int and ew
%*********************************************************
%*********************************************************

\begin{figure}
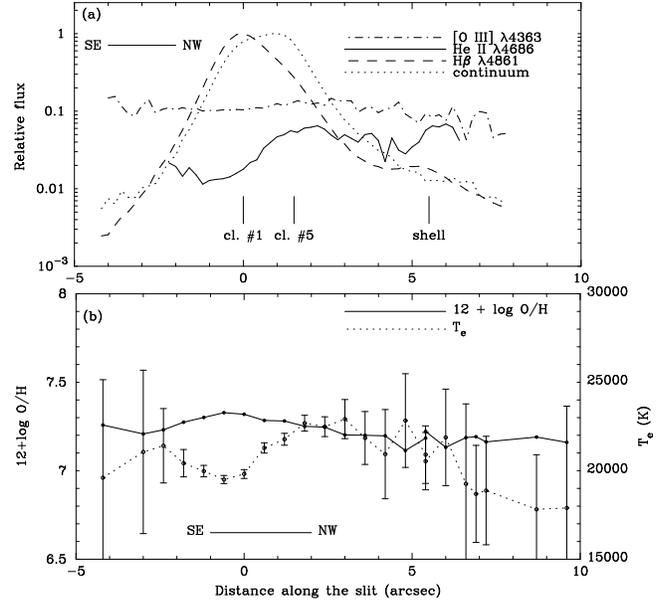
%[hbtp]
    \hspace*{0.0cm}\psfig{figure=Dh013_f2a.eps,angle=270,width=7.65cm}
    \hspace*{0.0cm}\psfig{figure=Dh013_f2b.eps,angle=270,width=8.5cm}
\caption{(a) The spatial distributions along the slit of
the relative fluxes of the emission lines and continuum
near H$\beta$. The fluxes of the H$\beta$ emission line and continuum are
normalized to their maximum values, while the fluxes of [ O {\sc iii}]
$\lambda$4363 and He {\sc ii} $\lambda$4686 are relative to the flux of
H$\beta$. The slit is oriented at position angle P.A. = -- 30$^\circ$. 
Origin is set at the maximum flux of the H$\beta$
emission line. (b) The spatial distribution of the electron temperature
$T_{\rm e}$ and oxygen abundance 12 + log O/H.}
    \label{fig:relint}
\end{figure}

%*********************************************************
%************************$*********************************

% =======================================
\section{Observations and data reduction}
% =======================================

The observations of SBS 0335--052 were carried out
on January 9, 2000 on the Keck II telescope with the low-resolution imaging
spectrograph (LRIS)
(Oke et al. 1995), using the
300 groove mm$^{-1}$ grating which provides a dispersion 
2.52 \AA\ pixel$^{-1}$ and a
spectral resolution of about 8 \AA\ in first order. 
The slit was 
1\arcsec$\times$180\arcsec, centered on the brightest H {\sc ii} region 
and oriented along the major axis with 
a position angle P.A. = --30$^{\circ}$. No binning along the spatial axis
has 
been done, yielding a spatial sampling of 0\farcs2 pixel$^{-1}$. The total 
exposure time was 60 min, broken into four
15 min exposures. All exposures were taken at an airmass of
1.1. The seeing was 0\farcs9.
%No correction for atmospheric dispersion was made because of 
%the small airmass during all observations. 
Wavelength calibration was provided by Hg-Ne-Ar comparison 
lamp spectra obtained after each exposure. Two spectrophotometric standard
stars, Feige 34 and HZ 44, were observed for flux calibration. 
The data reduction was made with the IRAF\footnote{IRAF is the Image 
Reduction and Analysis Facility distributed by the 
National Optical Astronomy Observatory, which is operated by the 
Association of Universities for Research in Astronomy (AURA) under 
cooperative agreement with the National Science Foundation (NSF).}
software package.
The two--dimensional spectra were bias--subtracted and 
flat--field corrected. Cosmic-ray removal, wavelength calibration, 
night sky background subtraction, correction for atmospheric extinction and 
absolute flux calibration were then performed.

For the analysis of spatial distribution of the brightest emission lines 
we also use the Keck spectrum of SBS 0335--052 obtained earlier by Izotov 
et al. (1999) at a position angle of 80$^\circ$. 

A distance of 54 Mpc to SBS 0335--052 is adopted throughout this paper
(Izotov et al. 1997). At this distance, 1\arcsec\ corresponds to 260 pc.

The spectrum of the brightest region in a 1\arcsec$\times$2\farcs2 aperture
with secure and tentative line identifications (see below) is
shown in Fig.\ 1. The measured S/N is of the order of 100.
Extinction-corrected fluxes of the emission lines normalized to the flux 
of 10$^{-4}$$\times$$I$(H$\beta$) are shown in Table~1.

%*********************************************************
%*********************************************************
%  Tab.1 - weak emission line intensities
%*********************************************************
%*********************************************************

\begin{table*}[tbh]
%     \centering{
\caption{Extinction-corrected fluxes of the emission lines in the brightest
region}
\label{t:Intens1}
\begin{tabular}{lrrcclrrcclrr} \hline \hline
$\lambda_{0}$(\AA) Ion &$\lambda_{\rm obs}$$^{\rm a}$&$I$($\lambda$)$^{\rm b}$&&&$\lambda_{0}$(\AA) Ion&$\lambda_{\rm obs}$$^{\rm a}$&$I$($\lambda$)$^{\rm b}$&&&$\lambda_{0}$(\AA) Ion&$\lambda_{\rm obs}$$^{\rm a}$
&$I$($\lambda$)$^{\rm b}$ \\ \hline
3727\ [O {\sc ii}]             &3728& 2136&&&4815\ [Fe {\sc ii}]               &4814&    9&&&5876\ He {\sc i}    &5875& 1043 \\
3750\ H12                      &3751&  393&&&4861\ H$\beta$                    &4861&10000&&&?                   &5917&    4 \\
3771\ H11                      &3771&  471&&&4907\ [Fe {\sc iv}]?              &4904&   22&&&?                   &5932&    3 \\
3798\ H10                      &3798&  597&&&4922\ He {\sc i}                  &4922&   94&&&5958\ Si {\sc ii}   &5957&    8 \\
3820\ He {\sc i}               &3820&   75&&&4959\ [O {\sc iii}]               &4959&10663&&&5979\ Si {\sc ii}   &5978&    8 \\
3835\ H9                       &3836&  798&&&5007\ [O {\sc iii}]               &5007&32271&&&?                   &5999&    3 \\
3868\ [Ne {\sc iii}]           &3869& 2349&&&5085\ [Fe {\sc iii}]              &5080&    6&&&6046\ O {\sc i}     &6044&    5 \\
3889\ H8\ +\ He {\sc i}        &3890& 1741&&&?                                 &5097&    3&&&6074\ He {\sc ii}   &6075&    3 \\
3926\ He {\sc i}               &3927&   27&&&5112\ [Fe {\sc ii}]               &5112&    6&&&?                   &6085&    2 \\
?                              &3943&   31&&&5146\ [Fe {\sc vi}]               &5149&   15&&&6102\ He {\sc ii}   &6100&    4 \\
3968\ [Ne {\sc iii}]\ +\ H7    &3969& 2478&&&5159\ [Fe {\sc ii}],[Fe {\sc vii}]&5160&   15&&&?                   &6150&    4 \\
4009\ He {\sc i}               &4009&   11&&&5176\ [Fe {\sc vi}]               &5181&    7&&&6170\ He {\sc ii}   &6169&    2 \\
4026\ He {\sc i}               &4027&  137&&&5199\ [N {\sc i}]                 &5198&   33&&&?                   &6270&    3 \\
4068--76\ [S {\sc ii}]         &4071&   44&&&?                                 &5214&    3&&&6300\ [O {\sc i}]   &6300&   87 \\
4101\ H$\delta$                &4102& 2604&&&?                                 &5235&    5&&&6312\ [S {\sc iii}] &6311&   51 \\
4121\ He {\sc i}               &4122&   16&&&?                                 &5244&    2&&&6347\ Si {\sc ii}   &6346&    8 \\
4144\ He {\sc i}               &4145&   20&&&5262\ [Fe {\sc ii}]               &5259&    7&&&6363\ [O {\sc i}]    &6363&  32 \\
4169\ He {\sc i}               &4168&   11&&&5271\ [Fe {\sc iii}]              &5270&  242&&&6407\ He {\sc ii}    &6407&   4 \\
4227\ [Fe {\sc v}]             &4228&   36&&&?                                 &5309&    6&&&?                   &6455&    9 \\
4287\ [Fe {\sc ii}]            &4288&   18&&&5335\ [Fe {\sc vi}]               &5336&    6&&&6563\ H$\alpha$     &6562&27366 \\
?                              &4314&   12&&&?                                 &5351&    2&&&6678\ He {\sc i}    &6677&  260 \\
4340\ H$\gamma$                &4341& 4753&&&?                                 &5369&    4&&&6717\ [S {\sc ii}]  &6716&  197 \\
4363\ [O {\sc iii}]            &4364& 1089&&&?                                 &5390&    2&&&6731\ [S {\sc ii}]  &6731&  166 \\
4388\ He {\sc i}               &4389&   43&&&5411\ He {\sc ii}                 &5412&   13&&&7002\ O {\sc i}     &7002&   13 \\
4415\ [Fe {\sc ii}]            &4416&   12&&&5424\ [Fe {\sc vi}]               &5424&    2&&&?                   &7038&    5 \\
?                              &4434&   10&&&?                                 &5436&    2&&&7065\ He {\sc i}    &7065&  427 \\
4452\ [Fe {\sc ii}]            &4453&    6&&&5485\ [Fe {\sc vi}]               &5482&    2&&&7136\ [Ar {\sc iii}]&7135&  212 \\
4471\ He {\sc i}               &4472&  344&&&?                                 &5501&    5&&&7171\ [Ar {\sc iv}] &7170&    7 \\
4511\ N {\sc iii}              &4509&    7&&&5517\ [Cl {\sc iii}]              &5517&   13&&&?                   &7195&    2 \\
4565\ Si {\sc iii}             &4566&   10&&&5538\ [Cl {\sc iii}]              &5537&   10&&&?                   &7210&    2 \\
?                              &4581&    8&&&?                                 &5552&    5&&&?                   &7224&    2 \\
4603-20\ N {\sc v}?,N {\sc iii}?&4609&  32&&&?                                 &5580&    3&&&7237\ [Ar {\sc iv}] &7237&    3 \\
4640\ N {\sc iii}              &4641&   13&&&?                                 &5595&    4&&&7254\ O {\sc i}     &7254&   10 \\
4658\ [Fe {\sc iii}]           &4659&   43&&&?                                 &5608&    3&&&7263\ [Ar {\sc iv}] &7264&    5 \\
4686\ He {\sc ii}              &4686&  221&&&5639\ [Fe {\sc vi}]               &5638&    3&&&7281\ He {\sc i}    &7280&   52 \\
4711\ [Ar {\sc iv}]+He {\sc i} &4712&  182&&&5677\ [Fe {\sc vi}]               &5678&    7&&&7298\ He {\sc i}    &7295&    5 \\
4740\ [Ar {\sc iv}]            &4741&  104&&&?                                 &5698&    2&&&7320\ [O {\sc ii}]  &7319&   60 \\
4755\ [Fe {\sc iii}]           &4757&    7&&&5721\ [Fe {\sc vii}]              &5721&   10&&&7330\ [O {\sc ii}]  &7330&   28 \\
\hline\hline
\end{tabular}

\noindent $^{\rm a}$$\lambda_{\rm obs}$ = $\lambda_{\rm measured}$/(1+$z$), 
where $\lambda_{\rm measured}$ is the measured wavelength of the
emission line, $z$ = 0.0135 is the redshift of SBS~0335--052.

\noindent $^{\rm b}$ $I$($\lambda$) is normalized to the flux of 
10$^{-4}$$\times$$I$(H$\beta$).
\end{table*}

%*********************************************************
%*********************************************************

%
% -----------------------------------------
\section{Weak nebular emission and stellar absorption lines}
% -----------------------------------------
%

%ds moved up
%In Fig. 1 we show the spectrum of the brightest region 
%(***aperture of***)
%of SBS 0335--052 with secure and tentative line identifications. 
%***The measured S/N is of the order of 100***.
%The extinction-corrected fluxes of the emission
%lines normalized to the flux of H$\beta$ are shown in Table 1.

One of main results (see Fig.\ 1) is the detection of numerous weak
permitted and 
forbidden emission lines, which are -- to the best of our knowledge --
detected for the first time in the spectrum of an extra-galactic giant
\hii\ region.
Lines with fluxes down to $\sim$ 0.1 \%\ of \hb\ are clearly detected.
For weaker lines the identifications are tentative and the measured
fluxes are fairly uncertain, with relative errors $\sim$ 100\%,
while the relative errors of the fluxes for the strongest lines are $\sim$ 1\%.
Thirty unidentified emission features with fluxes 0.02\% -- 0.12\% that 
of H$\beta$ are also shown in Table 1 and labeled by ``?''.

Some permitted lines, e.g. O {\sc i} $\lambda$6046, $\lambda$7002,
$\lambda$7254, Si {\sc ii} $\lambda$5958, $\lambda$5979, $\lambda$6347,
are likely of fluorescent origin and produced by absorption of the
intense UV radiation (Esteban et al. 1998). Other lines (broad
features at $\lambda$4620 and $\lambda$5700 -- 5850) can be stellar
in origin and likely produced by Wolf-Rayet stars. Weak 
He {\sc ii} $\lambda$4200 absorption is also detected, which is 
produced in the atmospheres of O stars.
Although our detection level is comparable to that in deep spectra
of Orion and planetary nebulae, optical recombination lines found in 
those objects (e.g.\ Esteban et al.\ 1998; Liu et al.\ 2000) are not 
detected here. This is likely due to the much lower metallicity of \sbs.

%Numerous weak permitted and forbidden emission lines are detected,
%which were never before seen in the spectra of the BCDs. 
%%The detection of weak lines in SBS 0335--052 opens opportunity for the
%%more detailed diagnostics of physical conditions in the ionized gas.
%Some permitted lines, e.g. O {\sc i} $\lambda$6046, $\lambda$7002,
%$\lambda$7254, Si {\sc ii} $\lambda$5958, $\lambda$5979, $\lambda$6347,
%are likely have fluorescent origin and produced by absorption of the
%intense UV radiation (Esteban et al. 1998). Some other lines (broad
%features at $\lambda$4620 and $\lambda$5700 -- 5850) can be stellar
%in origin and are produced by the Wolf-Rayet stars. The 
%He {\sc ii} $\lambda$4200 absorption line is detected, which is produced 
%in atmospheres of hot O stars.

Several emission lines indicate the presence of intense hard radiation at 
energies above 54 eV ($\lambda <$ 228 \AA) and possibly even above 75--99
eV.
The well-known He {\sc ii} $\lambda$4686 line and also 
He {\sc ii} $\lambda$5411 are detected.
We confirm the presence of the forbidden [Fe {\sc v}] $\lambda$4227
line, previously discovered in SBS 0335--052 and another low-metallicity 
BCD (Tol 1214--277) by Fricke et al. (2001).
In addition several weak [Fe {\sc vi}] and possibly [Fe {\sc vii}] 
emission lines are detected in our spectrum. 
The ionization potentials corresponding to these Fe lines are 
54.8, 75.0, and 99 eV respectively; the He$^+$ emitting region is
therefore expected to be associated with the emission
of [Fe~{\sc v-vii}]. The [Ar~{\sc v}] $\lambda$7006 line is likely detected.
However, it is blended with O~{\sc i} $\lambda$7002 emission line.
The possible identification of [Ar~{\sc v}] $\lambda$6435 line by Fricke 
et al.\ (2001) appears very uncertain from our spectrum.

%Strong He {\sc ii} $\lambda$4686 emission is present in the spectrum
%of SBS 0335--052. Additionally, the He {\sc ii} $\lambda$5411 
%emission line is detected. The presence of these lines indicates 
%intense hard radiation at wavelengths shorter 228\AA. We confirm
%the presence of the forbidden line [Fe {\sc v}] $\lambda$4227
%which is associated with the region of He {\sc ii} emission
%because the potential of ionization of Fe$^{3+}$ ion is 4.028 Ryd, i.e. 
%marginally higher than that of He$^+$.
%Previously this emission line was discovered by Fricke et al. (2001)
%in the spectra of SBS 0335--052 and another low-metallicity BCD
%Tol 1214--277. Besides that line several weak [Fe {\sc vi}] and
%possibly [Fe {\sc vii}] emission lines are detected in our spectrum. 

%*********************************************************
%*********************************************************
%  Fig.3 - Intensity distributions
%*********************************************************
%*********************************************************

\begin{figure}[hbtp]
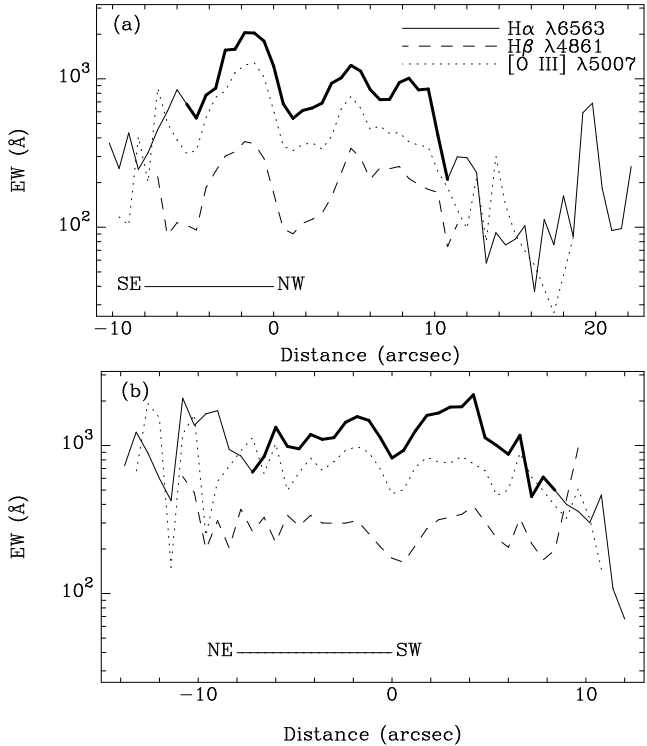

    \hspace*{0.0cm}\psfig{figure=Dh013_f3a.eps,angle=0,width=8.5cm}
    \hspace*{0.0cm}\psfig{figure=Dh013_f3b.eps,angle=0,width=8.5cm}
    \caption{(a) Distribution  along the slit of the equivalent widths 
of H$\alpha$ (solid line),
H$\beta$ (dashed line) and [O {\sc iii}] $\lambda$5007 (dotted line)
emission lines. The slit is oriented at position angle --30$^\circ$.
Thick solid line shows the range of the reliable $EW$(H$\alpha$). Outside
this range the continuum is very weak, corresponging to $R$ band surface 
brightness fainter 25 mag arcsec$^{-2}$.
(b) Same as (a) but slit is oriented at position angle 80$^\circ$.}
    \label{fig:intsect}
\end{figure}

The origin of hard radiation in giant H {\sc ii} regions has been debated
for many years. Possible sources for this radiation include shocks,
X-ray binaries, or hot Wolf-Rayet stars (cf.\ Garnett et al. 1991; Schaerer 
1996).
The spatial distribution of He {\sc ii} $\lambda$4686, [O {\sc iii}]
$\lambda$4363,
H$\beta$, and the adjacent continuum (see Fig.\ 2a) show some
evidence for the presence of shocks, at least in a limited area of \sbs.
Indeed, in the NW direction \Heii/\hb\ remains relatively 
strong\footnote{The significance of the dip at position 3--4 arcsec and the 
reincrease in \Heii/\hb\ to 6 arcsec is not large.}
out to the position of the ionized gas shell seen in HST images at a 
distance $\sim$ 6 arcsec from the two central clusters 1 and 5 of 
Thuan et al.\ (1997; cf.\ also Papaderos et al.\ 1998).
Although expected, the other high excitation lines of [Fe~{\sc v-vii}] are 
too faint to be detected away from the main clusters.
Furthermore, an increase of the electron temperature is observed away 
from the main clusters towards the shell (Fig.\ 2b), which may result
from additional heating due to shocks.
Both the extent of the region with \Heii/\hb\ $\sim$ 0.06 and the increase 
of $T_e$ at large distance from the main ionizing clusters, where geometric 
dilution will greatly reduce the local ionization parameter, are suggestive 
of shocks in the area within and out to the shell.
On the other hand, in the region centered on the main clusters we cannot 
distinguish between shocked and photoionized gas, as nearby stellar objects 
with hard ionizing spectra cannot be excluded.
In passing we note that if shocks are present, the oxygen abundance gradient
derived assuming a pure photoionized \hii\ region model (Fig.\ 2b) 
may not be real.
Observations of the spatial distribution of various high excitation
lines (including [Fe~{\sc v-vii}] and [Ne~{\sc v}] $\lambda$3426 which is 
expected to be strong from shock models; e.g.\ Dopita \& Sutherland 1996) 
as well as a detailed modeling are required to establish more firmly the 
importance of shocks on the spectrum of \sbs.

\section{Extended nebular emission}
%====================================================================

Melnick et al. (1992) first obtained an H$\alpha$ image of SBS 0335--052
and found that the ionized gas extends out to $\sim$~ 7 arcsec, or over a 
region of $\sim$ 2 kpc in diameter.
The deep long-slit Keck spectra of SBS 0335--052
allow us to trace ionized gas emission over a much larger region.
In Fig. 3 we show spatial distributions of the equivalent widths 
in two nearly perpendicular directions of the brightest emission lines 
H$\beta$, H$\alpha$
and [O {\sc iii}] $\lambda$5007. H$\alpha$ emission is detected
over 32 arcsec in the direction of P.A. = --30$^\circ$ and
over 26 arcsec in the direction of P.A. =  80$^\circ$.
This corresponds to a linear size of 8 $\times$ 6 kpc, roughly the size of
the 26 mag arcsec$^{-2}$ $R$ band isophote (Lipovetsky et al. 1999), 
or more than 3 times larger than that obtained from the H$\alpha$ image by 
Melnick et al. (1992). 

Very high equivalent widths of the emission lines in
SBS 0335--052 (Fig. 3) imply that the 
contribution of the ionized gas to the total light is important
and dominate in the regions with the largest equivalent widths.
This finding shows, as already pointed out in several earlier studies
(e.g.\ Kr\"uger et al. 1995; Izotov et al. 1997; Papaderos et al.\ 1998), 
that both nebular continuum
and line emission must be taken into account in photometric studies
of stellar populations in the extended regions of BCDs.
%It should be taken into account 
%in the analysis of the properties of stellar populations in the 
%extended regions of BCDs.
%We point out here the importance not only of the gaseous continuum
%emission but also the nebular line emission.
Obviously,
the contribution of the emission lines to the total light depends
on their relative strengths and redshift. 
H {\sc ii} region models predict a decrease of [O {\sc iii}]
$\lambda$5007/H$\beta$ 
with distance from the center due to a decreasing ionization parameter.
This is observed in SBS 0335--052. 
If one assumes that the extended emission in SBS 0335--052 is only gaseous, 
then the $V-I$ colour is changed from $\sim$ --0.6 mag in the center, 
where [O {\sc iii}] $\lambda$5007/H$\beta$ $\sim$ 3.3, to $\sim$ 0.0 mag
in outer regions, where [O {\sc iii}] $\lambda$5007/H$\beta$ $\sim$ 1.5.
In other words, even without stellar emission, the distribution
of gaseous emission mimics a contribution of red stars
increasing with distance. For analysis of stellar populations a proper 
removal of ionized gas emission is therefore crucial.

%The models of the photoionized
%H {\sc ii} regions predict that the [O {\sc iii}] $\lambda$5007/H$\beta$ 
%flux ratio is decreased from the center of the H {\sc ii} region outside 
%because of decreasing the ionization parameter. This is observed in
%SBS 0335--052. In particular, if one assumes that extended emission
%in SBS 0335--052 is only gaseous then the $V-I$ colour is changed
%from $\sim$ --0.6 mag in the center, where 
%[O {\sc iii}] $\lambda$5007/H$\beta$ $\sim$ 3.3, to $\sim$ 0.0 mag
%in outer regions, where [O {\sc iii}] $\lambda$5007/H$\beta$ $\sim$ 1.5.
%In other words, even in the case of no stellar emission the distribution
%of gaseous emission mimics the contribution of the red stars
%which increases with distance. Therefore, it is highly important in the
%analysis of stellar populations to subtract properly
%ionized gas emission.

%===========================================
%\section{Conclusion}
%===========================================

\begin{acknowledgements}
Y.I.I. has been partly supported by INTAS 97-0033 grant, and acknowledges
support from the ``Kiev project'' of the Universite Paul Sabatier of
Toulouse. 
He is grateful for the hospitality of the 
Midi-Pyrenees Observatory where this work was conducted.  The authors
further wish to extend special thanks to those of Hawaiian ancestry on whose
sacred mountain we are privileged to be guests.  Without their generous
hospitality, none of the observations presented herein would have been
possible.
\end{acknowledgements}

\end{document}